## Journal of Geophysical Research: Space Physics

# Interplanetary Magnetic Field $B_x$ Component Influence on Horizontal and Field-Aligned Currents in the Ionosphere

K. M. Laundal[1] [iD], J. P. Reistad[1] [iD], C. C. Finlay[2] [iD], N. Østgaard[1], P. Tenfjord[1] [iD], K. Snekvik[1] [iD], and A. Ohma[1] [iD]

[1]Birkeland Centre for Space Science, Department of Physics and Technology, University of Bergen, Bergen, Norway, [2]DTU Space, National Space Institute, Technical University of Denmark, Kongens Lyngby, Denmark

**Abstract** Statistical analyses have shown that the sunward component of the interplanetary magnetic field, $B_x$ (Geocentric Solar Magnetospheric), moderately but significantly affects the auroral intensity. These observations have been interpreted as signatures of a similar interplanetary magnetic field $B_x$ control on Birkeland currents yet to be observed directly. Such a control, attributed to differences in magnetic tension on newly opened magnetic field lines, would lead to stronger region 1 (R1) Birkeland currents for $B_x$ negative (positive) conditions in the Northern (Southern) Hemispheres than when $B_x$ is positive (negative). In this paper we perform a detailed investigation of three different sets of magnetic field measurements, from the Challenging Minisatellite Payload and *Swarm* low Earth orbit satellites, from the Active Magnetosphere and Planetary Electrodynamics Response Experiment products derived from the Iridium satellite constellation, and from the SuperMAG ground magnetometer network, each analyzed using different techniques, to test these predictions. The results show that a change in sign of $B_x$ changes the Birkeland currents by no more than $\approx 10\%$. The current patterns show little support for an interhemispheric asymmetry of the kind proposed to explain auroral observations. Instead, we propose an alternative interpretation, which is consistent with most of the auroral observations and with the current observations in the present paper, except for those based on Active Magnetosphere and Planetary Electrodynamics Response Experiment: The solar wind-magnetosphere coupling is more efficient when the dipole tilt angle and $B_x$ have the same sign than when they are different. We suggest that the higher coupling is because the dayside reconnection region is closer to the subsolar point when the dipole tilt angle and $B_x$ have the same sign.

**Plain Language Summary** The energy that powers geomagnetic activity and auroras originate in the solar wind. The most important parameters controlling the energy transfer is the solar wind velocity and the orientation of the magnetic field which is being carried by the solar wind. More energy will be transferred when this magnetic field is southward than when it is northward. Its component in the Earth-Sun direction is most often neglected, although some studies have shown that it has a modest influence on the intensity of auroras. In this paper we search for corresponding effects in electric currents in near-Earth space. We find that the sunward component of the magnetic field does seem to influence the currents, although not in the way suggested to explain the aurora observations. Instead, we propose that the sunward component of the interplanetary magnetic field modulates the rate at which energy is being transferred from the solar wind to the magnetosphere in a way which is different for different orientations of the Earth's magnetic field with respect to the Sun.

## 1. Introduction

The orientation of the interplanetary magnetic field (IMF) is the most important factor governing energy transfer from the solar wind to the terrestrial magnetosphere (e.g., Tenfjord & Østgaard, 2013). Among its three Geocentric Solar Magnetospheric (GSM) components, the $z$ component most strongly controls the energy transfer, since it largely determines whether or not the IMF and the Earth's magnetic field are antiparallel. When they are, magnetic reconnection takes place, and solar wind kinetic energy can be converted to magnetic energy in the magnetosphere. When the IMF has a strong $y$ component, which it usually does (Wilcox & Ness, 1965), magnetic tension on newly opened magnetic field lines diverts the subsequent flow of plasma differently in the two hemispheres, leading to significant asymmetries in the magnetosphere (Tenfjord et al., 2015) manifested in global convection patterns (e.g., Haaland et al., 2007; Heppner & Maynard, 1987; Pettigrew et al., 2010), currents (e.g., Anderson et al., 2008; Green et al., 2009; Laundal, Gjerloev, et al., 2016), and auroral intensity (e.g., Newell et al., 2004; Shue et al., 2001). The $x$ component, on the other hand, is most often assumed to be unimportant. It is the purpose of this paper to investigate whether this assumption is true when it comes to ionospheric currents. By ionospheric currents, we mean both the horizontal and the field-aligned currents in the ionosphere, mapped to 110-km altitude. The study is restricted to situations when subsolar reconnection dominates; during predominantly northward IMF, $B_x$ may have significant effects on lobe reconnection and associated dynamics (e.g., Østgaard et al., 2003; Shi et al., 2013), which we do not discuss here.

When the IMF has a strong component in the Earth-Sun ($x$) direction, newly opened magnetic field lines will have a very different curvature. That difference translates to a hemispheric asymmetry in magnetic tension forces which, according to Cowley (1981), result in an induced magnetic $x$ component in the magnetosphere with the same sign as the IMF $B_x$. Although this process is analogous to how $B_y$ leads to hemispheric asymmetries, it is ostensibly of much less consequence for global geospace. We are only aware of few studies that investigate the global effects of IMF $B_x$ observationally. One focuses on convection (Förster et al., 2011) and the others on auroral intensities: Shue et al. (2002) reported global average patterns of auroral ultraviolet (UV) luminosity, observed by Polar Ultraviolet Imager from above the Northern Hemisphere, showing stronger aurora when $B_x < 0$ than when $B_x > 0$ (when $B_z < 0$). Based on Polar Ultraviolet Imager observations from January 1997, Baker et al. (2003) reached similar conclusions. Reistad et al. (2014), motivated by observations from single events showing nonconjugate aurora during strong $B_x$ conditions (Laundal & Østgaard, 2009; Reistad et al., 2013), did a statistical analysis of UV auroral images from the Wideband Imaging Camera on the Imager for Magnetopause-to-Aurora Global Exploration (IMAGE) satellite, in search of $B_x$ effects. Their observations, which focused on the dusk region, showed a weak but statistically significant anticorrelation between $B_x$ and auroral intensity in the north and a correlation in the south. Both Shue et al. (2002) and Reistad et al. (2014) explained their findings in terms of corresponding asymmetries in upward field-aligned currents, which are carried in part by the downward electrons that excite auroras. In terms of the average picture first discovered by Iijima and Potemra (1978), these upward currents are the region 1 current at dusk and region 2 current at dawn, the former being the strongest.

Even though Shue et al. (2002) and Reistad et al. (2014) observed the same $B_x$ dependence and explained their findings in terms of current asymmetries inferred from the Cowley (1981) picture, their arguments for how this picture implies asymmetric currents differed. According to Reistad et al. (2014), the geometry suggested by Cowley (1981) on newly opened field lines during strong IMF $B_x$ implies different magnetopause currents above the polar regions. Their conjecture was that this asymmetry also affects the region 1 currents, which in turn controls the auroral intensity. The effect of different field line curvature presumably decreases as the open field lines travel far downtail, consistent with observations of the strongest asymmetries at magnetic local times earlier than 20. The effect should be present also in the dawn region 1 current, although this is not manifested in electron auroras. It is not clear how or if differences in magnetopause currents affect region 2 currents. Shue et al. (2002), on the other hand, argued from a purely geometrical point of view: An induced $B_x$ in the magnetosphere will shift the footpoints of magnetospheric field lines toward or away from the Sun, the direction depending on $B_x$ and hemisphere. This shift was thought to increase or decrease ionospheric flow shears, with corresponding changes in Birkeland currents. As a consequence, Shue et al. (2002) predict stronger R1 currents on the nightside in the Northern Hemisphere when $B_x$ is negative, similar to Reistad et al. (2014), but they predict the opposite asymmetry in the dayside part of the R1 current. Furthermore, they predict a $B_x$-dependent asymmetry in the R2 current which is similar to the asymmetry in the adjacent R1 current.

To our knowledge, no direct observations of currents have been presented to test these predictions. This is what we undertake in this paper. Since the $B_x$ effects, if they exist, are weak, it is an advantage to consider multiple tests using different data sets. In the next section we start by presenting the results from an empirical model of ionospheric currents, specifically designed to reveal asymmetries associated with IMF $B_x$, based on magnetic field observations from the Challenging Minisatellite Payload (CHAMP) and Swarm satellites (section 2.1). Then we present statistics based on the Active Magnetosphere and Planetary Electrodynamics Response Experiment (AMPERE), which offers global Birkeland current maps based on magnetometers on the Iridium satellites (section 2.2). Finally, we present global maps of the equivalent horizontal current, calculated using ground magnetometers (section 2.3). In section 2.4 we present a summary of all the tests. In section 2 we focus primarily on the predictions made by Reistad et al. (2014) but return to discuss our results in terms of the Shue et al. (2002) predictions in section 2.4. In section 3 we discuss the implications of our findings, and section 4 concludes the paper.

In order to isolate the effect of the IMF $B_x$, a number of biases must be considered. The IMF lines typically form a spiral, called the Parker spiral (Parker, 1958). Parker spiral IMF vectors at 1 AU are either toward the Sun and along Earth's orbit or away from the Sun and opposite to Earth's orbit. This means that the GSM $B_x$ and $B_y$ components are significantly correlated; we calculate a correlation coefficient of $-0.39$ using $> 12 \cdot 10^6$ 1-min OMNI data points from 1981 to 2016. In addition, there is a seasonally dependent correlation of the IMF $B_y$ (and thus $B_x$) with geomagnetic activity: Near spring equinox, geomagnetic activity is larger when $B_y$ is negative than when it is positive. In the fall, the situation reverses. This was demonstrated by Zhao and Zong (2012). They explained the annual variation in correlation between $B_y$ and geomagnetic activity as an aspect of the more fundamental Russell-McPherron (R-M) effect (Russell & McPherron, 1973). The R-M effect is a semiannual and diurnal variation in geomagnetic activity due to variations in the orientation of Earth's magnetic field with respect to the Parker spiral IMF. The basic idea is that the Parker spiral magnetic field is in the $xy$ plane in Geocentric Solar Equatorial coordinates and that its projection on the $yz$ plane in GSM coordinates changes. It should be noted, considering the topic of the present paper, that the $x$ component is the same in Geocentric Solar Equatorial and GSM coordinates and is not directly part of the R-M effect. Nevertheless, the Parker spiral orientation and the seasonal variations of the R-M effect mean that the most important controlling parameters for ionospheric currents are strongly correlated. Because of these correlations between controlling parameters, we severely constrain the data selections in the observations presented below, in order to minimize the contribution from other parameters, in particular the IMF $B_y$. Possible biases are also considered in detail in Appendix A. In Appendix B, we present the same figures as in section 2, except that the roles of $B_x$ and $B_y$ are reversed. The purpose of this is to demonstrate that our techniques are capable of reproducing known variations in currents with $B_y$. That indicates that they would also be capable of isolating $B_x$ effects, if these exist.

## 2. Observations

In this section we present tests of the $B_x$ effect using three different data sets. For each data set, we present global average patterns of Birkeland currents and/or equivalent currents, but the techniques used to calculate the averages differ. We use only data from periods when $B_z$ was negative, which implies favorable conditions for subsolar reconnection and less so for lobe reconnection. Also, each data set is split into two disjoint sets, defined by the dipole tilt angle being either $< -10°$ or $> 10°$. This is done in order to reveal potential seasonal dependencies (the results by Reistad et al., 2014 were based only on winter observations) and also to increase the number of tests based on statistically independent data points. Each set is used to provide independent comparisons between corresponding average currents for different signs of $B_x$. Apart from visually comparing the morphology of the current, we use peak current values as a metric for quantitative comparison of Birkeland currents.

### 2.1. Test 1: CHAMP and Swarm

Thanks to the CHAMP and Swarm satellite missions, we now have access to many years of incredibly precise measurements of the magnetic field at low Earth orbit, providing global coverage for studies of a large range of external conditions. This data set is therefore an ideal basis for climatological models of global ionospheric currents. Here we present results from two such models, parametrized in terms of the orientation of the IMF. The two models are based on data from periods when then dipole tilt was either $> 10°$ or $< -10°$, respectively. The data were sampled at 30-s cadence, and a high-resolution model of the near-Earth (main) geomagnetic field, CHAOS-6 (Finlay et al., 2016), was subtracted. CHAOS-6 includes estimates of the core field, the lithospheric field, and the quiet time near-Earth field due to magnetospheric currents.

The remaining field, $\Delta\mathbf{B}$, is a sum of poloidal and toroidal parts, which can be written in terms of scalar fields $V$ and $T$, respectively (e.g., Backus, 1986; Olsen, 1997):

$$\Delta\mathbf{B} = -\nabla V + \mathbf{r} \times \nabla T, \qquad (1)$$

where $\mathbf{r}$ is a radial vector. $V$ relates to currents that do not intersect the satellite orbits (including divergence-free horizontal currents in the ionosphere), and $T$ relates to currents that cross the orbits (i.e., the field-aligned currents). We express $V$ and $T$ in terms of spherical harmonics, defined in apex quasi-dipole and modified apex coordinates (Richmond, 1995a), respectively, with magnetic local time (Laundal & Richmond, 2017) as the azimuthal coordinate. We take the nonorthogonality of these coordinate systems into account when evaluating the gradients in (1). The technique is described in detail and demonstrated by Laundal, Finlay, and Olsen (2016), who applied it to data sets binned by solar wind and seasonal conditions. We refer to that paper for a detailed mathematical description. Here we extend the technique by allowing the spherical harmonic coefficients to vary as a function of the IMF and solar wind velocity.

The nature of the relationship between spherical harmonic coefficients and these external parameters is not trivial. Weimer (2013) found that, for a general model of ground magnetic field perturbations, it was beneficial to use a Fourier expansion in IMF clock angle $\theta_c$ (the angle between the IMF in the $yz$ plane and the GSM $z$ axis), scaled by various terms thought to influence the currents (tilt angle, solar wind velocity, IMF magnitude, and $F_{10.7}$). Inspired by this, we choose a similar function, adapted for our purposes. Each spherical harmonic coefficient in the expansion of $V$ and $T$ (see equations (7) and (9) of Laundal, Finlay, & Olsen, 2016), say $g_n^m$, is expanded as

$$g_n^m = a_n^m + b_n^m \epsilon + c_n^m \epsilon \cos(\phi) + d_n^m \epsilon \sin(\phi) + e_n^m \epsilon \cos(2\phi) + f_n^m \epsilon \sin(2\phi). \qquad (2)$$

This is a Fourier expansion in IMF azimuth angle, $\phi$, scaled by a solar wind coupling function, $\epsilon$. The azimuth angle is defined as the angle between the IMF in the $xy$ plane and the GSM $x$ axis. The $\epsilon$ parameter used here is that proposed by Newell et al. (2007), $\epsilon = |v_x|^{4/3} B_{yz}^{2/3} \sin^{8/3}(\theta_c/2)$, where $v_x$ is the solar wind velocity in the GSM $x$ direction and $B_{yz}$ is $\sqrt{B_y^2 + B_z^2}$. The numerical values of $\epsilon$ reported in this paper are calculated with $v_x$ in units of kilometer per second, $B_{yz}$ in units of nanotesla, and then scaled by $10^{-3}$. The key property of $\epsilon$ is that it presumably correlates well with dayside reconnection rate, although its unit is physically meaningless and therefore not reported. We only use data from periods when $B_z < 0$. The solar wind and IMF parameters are determined from 1-min OMNI data, averaged over the 20 min prior to the time of the corresponding Swarm/CHAMP measurement. In total, we use 8,211,033 measured vector components to determine the coefficients of the model for negative tilt angles ($< -10°$) and 9,195,507 components for the positive tilt angle model ($> 10°$). The CHAMP measurements are from August 2000 to September 2010 and Swarm from December 2013 to August 2016.

The spherical harmonic series are truncated at $n, m = 35, 5$ for $V$ and at $n, m = 60, 5$ for $T$. That leads to 1,005 real coefficients which are each expanded in terms of the six parameters of equation (2). This equation, together with equations (10), (11), and (12) in Laundal, Finlay, & Olsen (2016), relates a vector of measurements, $\mathbf{d}$ to a vector of 6,030 model coefficients, $\mathbf{m}$. The full set of equations can be formulated as a matrix equation, $\mathbf{Gm} = \mathbf{d}$, where $\mathbf{m}$ can be estimated using an iteratively reweighted least squares scheme that robustly handles long-tailed distributions of data errors. In the $(i + 1)$th iteration, $\mathbf{m}_{i+1}$ is determined by

$$\mathbf{m}_{i+1} = (\mathbf{G}^T \mathbf{W}_i \mathbf{G} + \alpha \mathbf{R})^{-1} \mathbf{G}^T \mathbf{W}_i \mathbf{d}, \qquad (3)$$

where $\mathbf{W}_i$ is a weight matrix that depends on the misfit in the $i$th iteration, $\mathbf{e}_i = \mathbf{Gm}_i - \mathbf{d}$. The diagonal elements of $\mathbf{W}_i$ are called Huber weights (Huber, 1964) and are defined as

$$W_{jj} = \min(1, 1.5\sigma/|e_i^j|), \qquad (4)$$

where $e_i^j$ is the $j$th element of $\mathbf{e}_i$ (a vector with as many elements as there are measurements). $\sigma$ is in this case determined as the square root of the Huber-weighted mean squared element of $\mathbf{e}_i$. In addition to the variable weights, we weight observations from the side-by-side flying satellites, Swarm Alpha, and Charlie, by 0.5, since they are assumed not to provide independent information on the length scales considered by our model.

The inversion in equation (3) is done by the use of Cholesky decomposition and the Python Scipy function cho_solve, which makes use of the LAPACK library (Anderson et al., 1999). This technique fails unless we apply some regularization for the toroidal field. The reason for that is probably that the representation in modified apex coordinates is not appropriate for low latitudes (Matsuo et al., 2015), and a lack of coverage near the modified magnetic apex equator, which is at the dip equator at the chosen reference height, 110 km (see Richmond, 1995a, for details). Therefore, the regularization matrix $\mathbf{R}$ in equation (3) is chosen to measure the mean square vector toroidal field due to the field-aligned currents (see, e.g., Sabaka et al., 2010, equation 116) and is a diagonal matrix, with elements $\alpha n(n + 1)/(2n + 1)$ in the columns that correspond to the elements of **m** that are coefficients in the expansion of $T$, and zero otherwise. The $\alpha$ was set to 100, the smallest power of 10 that prevented cho_solve from failing. No regularization is used for $V$. The iterations were terminated when $\|\mathbf{m}_i - \mathbf{m}_{i+1}\| < 0.04\|\mathbf{m}_0\|$. This occurred after seven iterations in the negative tilt angle model and six iterations for the positive tilt angle model.

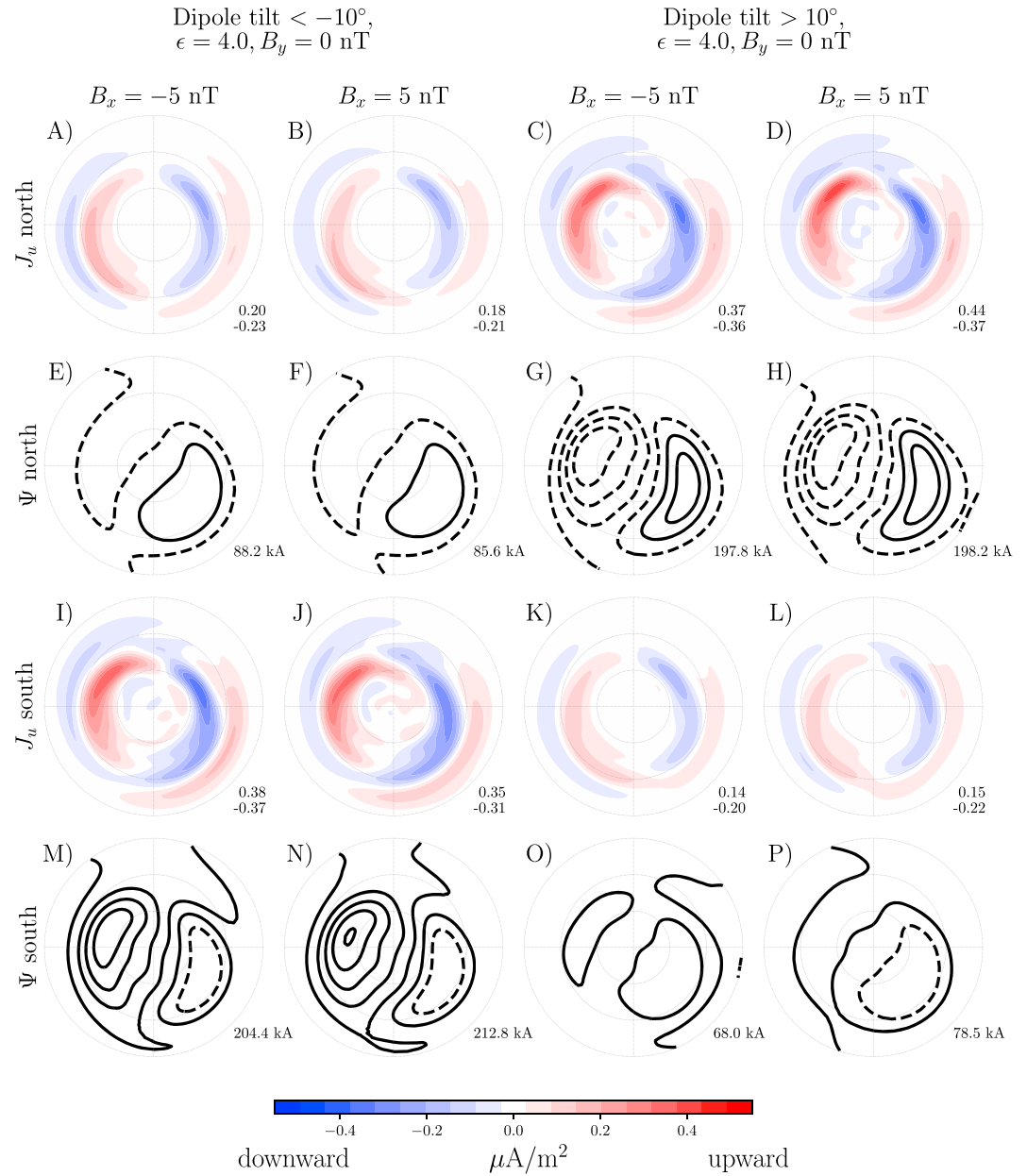

**Figure 1.** Birkeland currents (color) and horizontal equivalent currents (black contours) from two different empirical models of ionospheric currents, based on magnetic field measurements from the CHAMP and Swarm satellites. (a, e, i, m and b, f, j, n) Based on data from periods when the dipole tilt angle was < −10°. (c, g, k, o and d, h, l, p) Data from periods when the tilt angle was >10°. The models depend on both $B_x$ and $B_y$, and in this figure we set $B_y = 0$ to highlight any $B_x$ effect. The Newell et al. (2007) coupling function $\epsilon$ is set to 4. Numbers in the lower right corner of the Birkeland current plots indicate peak upward (positive) and downward (negative) currents, in $\mu$A/m². The number in the lower right corner of the equivalent current plots indicates the total current flowing between the maximum and minimum of the current function. An equivalent current of 30 kA flows between each contour.

The iterative scheme progressively downweights outlying data points that deviate strongly from model values. This procedure reduces the impact of such outliers and enables the final solution to better represent typical values rather than simple mean values. The misfit in the final iterations, as quantified as the root Huber-weighted mean square element of **e**, was 17.2 nT in the negative tilt angle model and 16.2 nT in the positive tilt angle model. The root mean squares of the data points for the corresponding models were 62.1 and 57.1 nT, respectively. The mean Huber weight in the final iteration was 0.87 for both models.

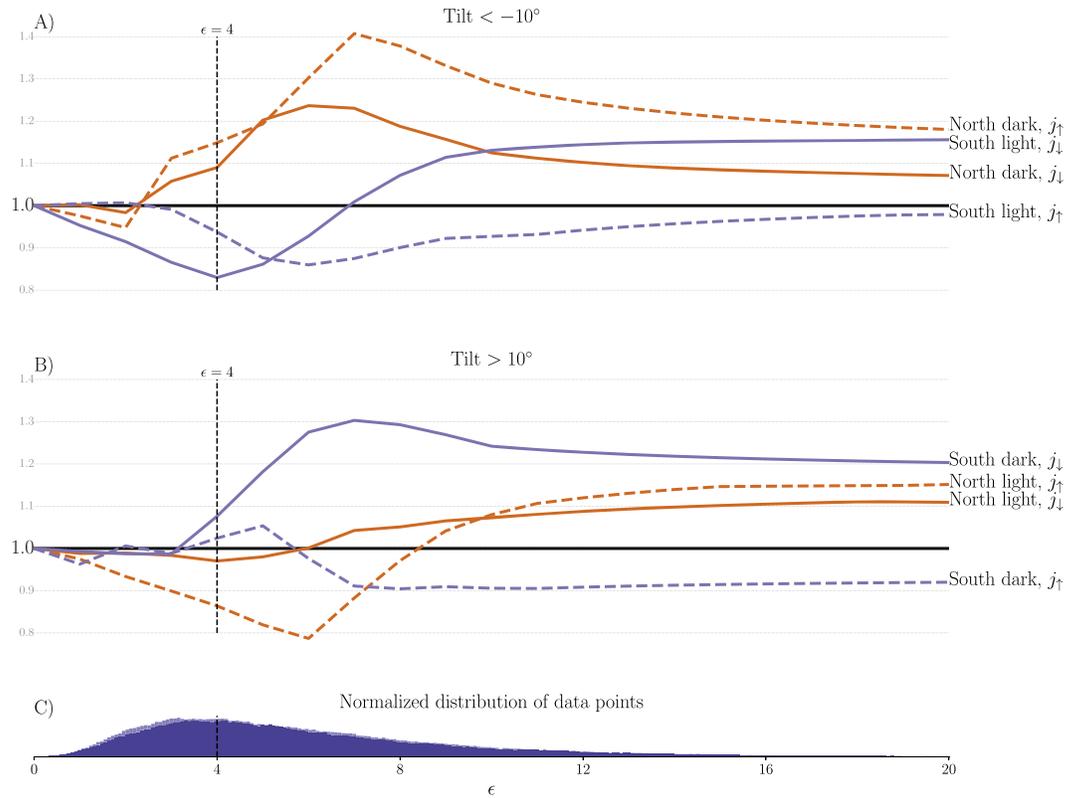

**Figure 2.** Ratios of corresponding R1 currents for different signs of $B_x$, as a function of $\epsilon$. The ratios are defined such that they are > 1 if the asymmetry is consistent with the Reistad et al. (2014) predictions. The curves for the two models are shown separately in (a) (tilt < −10°) and (b) (tilt > 10°). The normalized distribution of $\epsilon$ is shown in (c). The distributions for the two models overlap where the color is dark blue.

The curl of equation (1) gives the currents associated with $\Delta \mathbf{B}$. From $T$ we can analytically calculate field-aligned currents at some height (we use 110 km). From $V$ we can calculate the divergence-free part of a horizontal current sheet under the satellite orbits, also assuming a height of 110 km. We represent the divergence-free sheet current density $\mathbf{j}_{df}$ in terms of a scalar current function $\Phi$: $\mathbf{j}_{df} = \mathbf{k} \times \nabla \Phi$, where $\mathbf{k}$ is an upward unit vector. The divergence-free current function is very similar to the equivalent current function derived with ground magnetometers (we return to this in section 2.3). Equations relating the spherical harmonic coefficients in the expansion of $\Delta \mathbf{B}$ to currents can be found in Laundal, Finlay, and Olsen (2016; see their equations (15) and (16)).

Figure 1 shows Birkeland currents and equivalent currents in the Northern and Southern Hemispheres from the two models for positive and negative tilt angles. To produce this figure, we have evaluated the model setting $\epsilon = 4$, which is close to the peak of the distribution of $\epsilon$ in the two data sets. We also set $B_y = 0$, to highlight differences related to $B_x$. $\epsilon = 4$ and $B_y = 0$ could correspond to a solar wind speed of about 290 m/s, and $B_z = −3$ nT. The negative dipole tilt angle model is shown in the two columns to the left, and the positive dipole tilt angle model is shown in the columns to the right. The two upper rows represent Northern Hemisphere patterns, and the two lower rows represent Southern Hemisphere patterns.

The $B_x$ effects should appear as differences between horizontal image pairs in the two model columns. It is immediately clear from visual inspection that any such differences are small. This is in contrast to the IMF $B_y$, which has a pronounced influence on current morphologies (Anderson et al., 2008; Green et al., 2009; Laundal, Gjerloev, et al., 2016; Weimer, 2001).

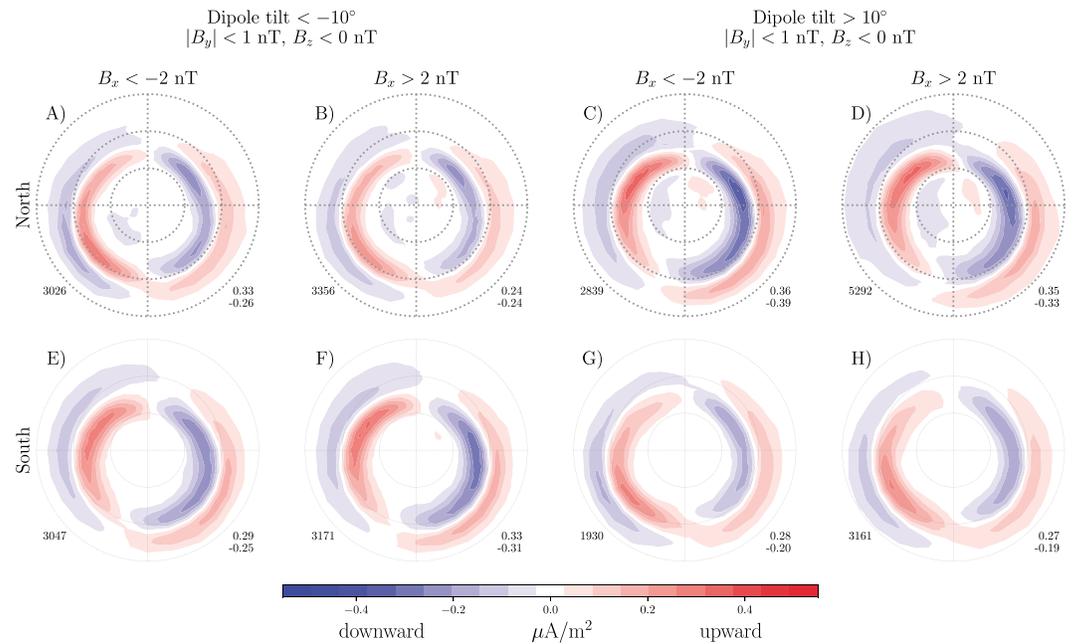

**Figure 3.** Maps of average AMPERE Birkeland currents from both hemispheres during $B_z$ negative/$B_x$-dominated conditions. The Northern Hemisphere is shown in the top row and the Southern Hemisphere in the bottom row. The plots in (a, e, b, f) are based on data from periods when the tilt angle was $< -10°$, and (c, g, d, h) on periods when the tilt was $> 10°$.

The numbers in the lower right corners of the Birkeland current plots in Figure 1 correspond to the peak current values, in microampere per square meter. The maxima consistently correspond to peak upward R1 currents (dusk), and the minima correspond to peak downward R1 currents (dawn). Comparing peaks between the horizontal pairs, we find that four out of eight are slightly asymmetrical in the direction predicted by Reistad et al. (2014; stronger R1 currents for $B_x < 0$ in the north and opposite in the south), while the remaining pairs show the reverse. All differences are, however, rather small, being only about 10%. We note that all winter hemisphere asymmetries are consistent with the Reistad et al. (2014) predictions, while the summer hemisphere results are not. We return to this seasonal difference in $B_x$ dependence in section 3.

Figure 1 only shows the asymmetries for cases when $\epsilon = 4$. The arguments presented by Reistad et al. (2014) suggest that stronger solar wind velocity, and stronger magnetic field, will enhance the asymmetry. In Figure 2 we test if the asymmetries become clearer when $\epsilon$ increases. The figure presents the ratios of corresponding R1 currents for different signs of $B_x$ as a function of $\epsilon$. The ratios are defined such that they are greater than 1 if the Reistad et al. (2014) predictions are fulfilled. We see that as $\epsilon$ becomes very large, six out of eight ratios are $> 1$, supporting the predictions. However, we emphasize that large values for $\epsilon$ are rare in the data set used to make the model, as shown in the normalized distributions in Figure 2c, and the models presumably become similarly more uncertain in this extreme regime.

In addition to the Birkeland currents, Figure 1 also shows equivalent ionospheric currents. It is not obvious how the $B_x$ effect would be expected to affect equivalent current; except that when $B_x$ favors strong R1 currents the equivalent currents should be stronger too. From visual inspection, there is no appreciable difference in the estimated equivalent currents between $B_x$ positive and negative conditions. However, in all but one case, the total current, given in the lower right corners, is slightly stronger when the Reistad et al. (2014) predictions indicate stronger currents. The only case where this is not true is in the Northern Hemisphere for tilt $> 10°$ (Figures 1g and 1h), where the difference is less than 1%.

We have chosen not to focus on hemispheric differences in Figure 1; even though in principle we could compare the negative (positive) tilt model in the Northern Hemisphere to the positive (negative) tilt model in the south to look for interhemispheric $B_x$ effects. The reason for this is that there are inherent asymmetries between hemispheres that could easily contribute differences of similar magnitudes as the effect of $B_x$. The most important hemispheric asymmetries are different offsets between geographic and magnetic poles and different field strengths. Differences in pole offset lead to different variations in insolation, and thus conductivity, in the two hemispheres, which are not fully described by the dipole tilt angle (Laundal, Gjerloev, et al., 2016). Differences in magnetic field strength also lead to differences in ionospheric conductivity directly through its influence on ionospheric plasma (Cnossen et al., 2012; Richmond, 1995b) and indirectly through variations in the particle mirror height, changing the flux of ionizing particle precipitation (Stenbaek-Nielsen et al., 1973). These differences may remain important even after compensating for geometric differences in the main field which we do here by use of magnetic apex coordinates. See Laundal et al. (2017) for a review of how north-south asymmetries in the main field may affect geospace phenomena.

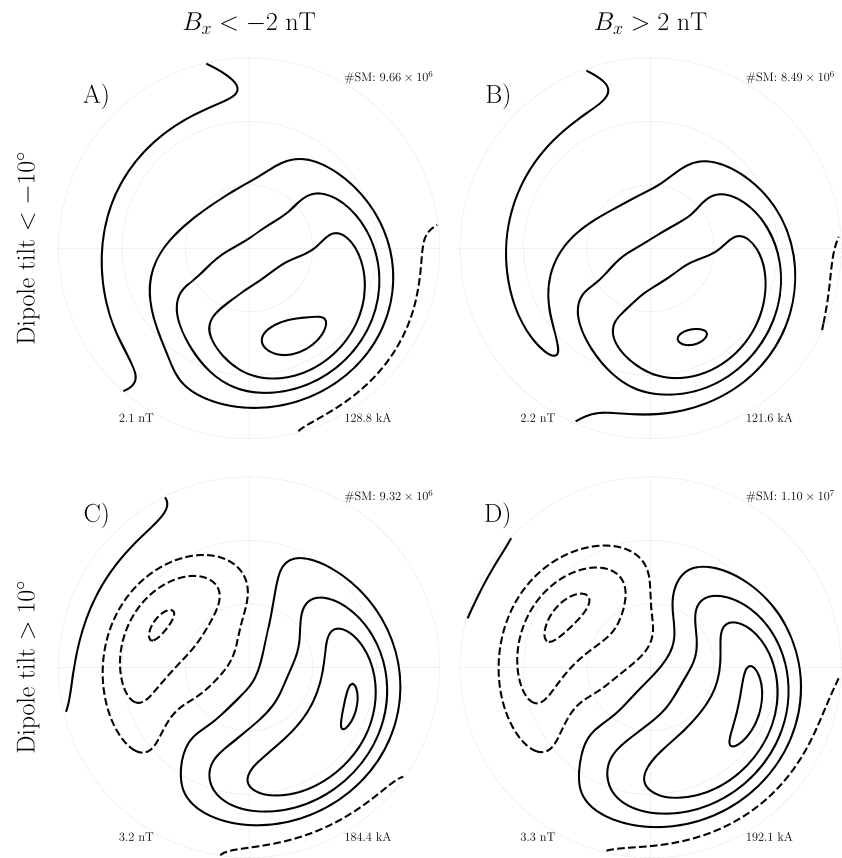

**Figure 4.** Equivalent current patterns derived from Northern Hemisphere ground-based magnetometer measurements in the period 1981–2014, obtained when $B_z < -1$ and $|B_y| < 1$ nT. Interplanetary magnetic field $B_x > 2$ in (a, c) and $B_x < -2$ nT in (b, d). In (a, b), the dipole tilt angle is $> 10°$ (summer) and in (c, d) is $< -10°$ (winter). A full description of the technique is given in Laundal, Gjerloev, et al. (2016). An equivalent current of 30 kA flows between each contour.

## 2.2. Test 2: AMPERE Maps of Field-Aligned Currents

Figure 3 shows global average Birkeland currents, calculated using AMPERE maps between January 2010 and May 2013. AMPERE provides polar maps of field-aligned currents at 2-min cadence and 10-min integration time, based on magnetometer measurements from the fleet of commercial Iridium satellites and spherical harmonic analysis (Anderson et al., 2000; Waters et al., 2001). The plots in Figure 3 differ by data selection criteria: The top row is based on data from the Northern Hemisphere and the bottom row from the Southern Hemisphere. The two columns to the left are based on data from periods when the tilt angle was $< -10°$ and the right plots on periods when the tilt angle was $> 10°$. In addition, the columns correspond to different signs of $B_x$, as indicated above the top row plots. For all the plots, $B_z < 0$, $|B_x| > 2$, and $|B_y| < 1$ nT, ensuring strong solar wind coupling and little contamination from $B_y$ effects. Each map was also subject to a stability selection criterion, based on the similarity with the map from 20 min earlier. We use the same similarity metric as Anderson et al. (2008), the fractional overlap, which we require to be $> 0.45$.

The final number of maps available for calculating the average is given in the lower left corner of the plots. The average was calculated in a robust way, using Huber weights, largely analogous to what was done for the model estimation in the previous section: We start by calculating the simple mean map and standard deviation map. Then, for each map, we make a map of weights, which are 1 if the distance from the mean is less than 1.5 in units of the standard deviation, and decreasing in inverse proportion to that distance for larger deviations. These weights are used in a new iteration to calculate an updated weighted mean. The corresponding weights are then also updated and new mean maps again calculated until the difference from the previous iteration is small. This robust procedure helps to reduce the impact of outliers.

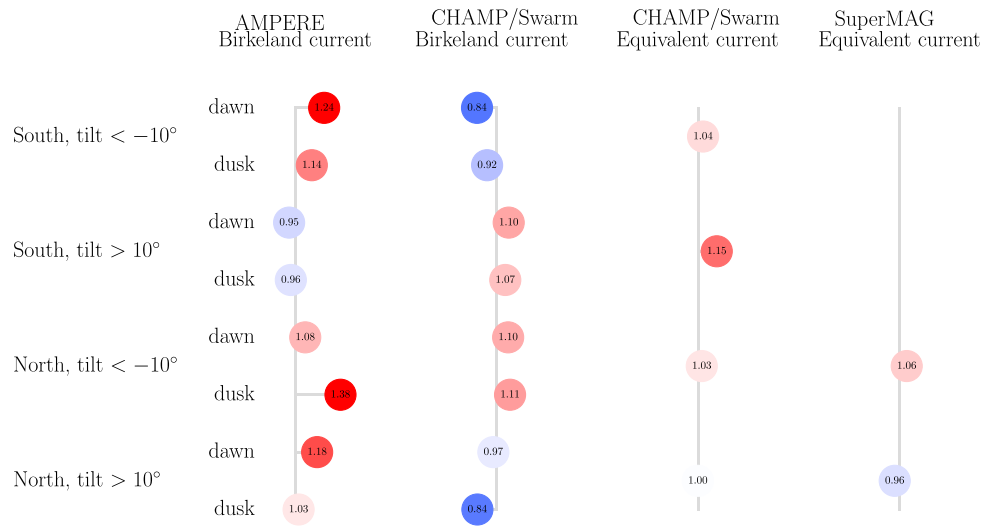

**Figure 5.** The ratios of corresponding currents during presumably favorable interplanetary magnetic field $B_x$ conditions to unfavorable conditions (numbers > 1 indicate that the asymmetry is consistent with the predictions by Reistad et al., 2014). Birkeland current ratios are based on the peak values listed in Figures 1 and 3. For the equivalent currents, the ratios are based on the total current, listed in Figures 1 and 4. AMPERE = Active Magnetosphere and Planetary Electrodynamics Response Experiment; CHAMP = Challenging Minisatellite Payload.

Visual inspection of Figure 3 reveals no clear difference between different signs of $B_x$. Investigating the peak R1 values, we find that six out of eight have an asymmetry in the direction predicted by Reistad et al. (2014). Among those are the largest difference found in this study, a 38% stronger peak R1 current at northern dusk when the tilt angle is < −10°. Thus, the AMPERE results are slightly more supportive of the Reistad et al. (2014) predictions than the analysis based on Swarm and CHAMP data, although the interpretation of the results is not without ambiguity.

### 2.3. Test 3: SuperMAG Ground Magnetometers: Equivalent Currents From Binned Averages

As a final test of the $B_x$ influence, we present equivalent current patterns based on ground magnetometer measurements. The ground magnetometer measurements are processed and made available by SuperMAG (Gjerloev, 2012). The SuperMAG measurements were obtained in the years 1981–2014; converted to quasi-dipole coordinates (Laundal & Gjerloev, 2014; Laundal, Gjerloev, et al., 2016), binned in 920 grid cells; sorted according to the IMF orientation, $B_z < −1$ nT, $|B_y| < 1$ nT, $B_x$, and tilt angle > 10° or < −10°; and averaged taking the simple mean. The average vectors were then fitted with a spherical harmonic representation of external and internal magnetic potentials as described by Laundal, Gjerloev, et al. (2016). The external magnetic potential corresponds to ionospheric currents, but it cannot be used to derive the full three-dimensional current system without any additional information. Instead, we use it to calculate an equivalent current, a sheet current in a spherical shell at 110 km that would produce a magnetic field that is equal to the observed external field. We represent the equivalent current in terms of a scalar $\Psi$, analogous to the divergence-free current function derived from space: $\mathbf{j}_{eq} = \mathbf{k} \times \nabla \Psi$, where $\mathbf{j}_{eq}$ is the horizontal equivalent sheet current density. We refer to Laundal, Gjerloev, et al. (2016) for a detailed description of how $\Psi$ is estimated and to Laundal, Finlay, and Olsen (2016) for a discussion about the differences between the equivalent current function $\Psi$ and the current function $\Phi$, derived with satellite magnetometers. Figure 4 shows contour plots of the equivalent current function associated with the external magnetic potential. The total current flowing between the maximum and minimum of the current function is given in the lower right corners. The root mean square error of the spherical harmonic fit is given in the lower left corners. The top right corners indicate the total number of SuperMAG measurements that were used to make each plot. Only patterns from the Northern Hemisphere are shown, since the ground magnetometer coverage is much better there.

Visual inspection reveals little difference between currents for different signs of $B_x$. The difference in total current is in the order of 5%. The difference is consistent with the Reistad et al. (2014) predictions during negative tilt but not during positive tilt.

### 2.4. Summary of Results

We have presented maps of ionospheric currents from three different data sources, using three different techniques. The maps offer several opportunities to compare currents in search of differences related to the IMF $B_x$ component. We can conclude already, based on visual inspection of the maps, that any $B_x$ influence must be small, of the order of $\approx 10\%$.

We have used the ratios between corresponding peak currents as a quantitative metric of the $B_x$ associated asymmetries. The ratios are not unambiguously in the direction suggested by the Reistad et al. (2014) predictions. Figure 5 shows all the ratios, visualized as increasingly red or blue dots, depending on the sign of the differences. Red dots indicate asymmetries that are consistent with Reistad et al.'s (2014) predictions for the $B_x$ effect. There are slightly more red dots than blue dots, but there is no evidence for a very strong influence of $B_x$.

So far we have focused on the $B_x$ predictions made by Reistad et al. (2014). However, as mentioned in section 1, Shue et al. (2002) interpret the Cowley (1981) picture somewhat differently, leading to a different set of predictions for how $B_x$ affects field-aligned currents: In the Northern Hemisphere, both the R1 and R2 currents should be stronger on the nightside when $B_x < 0$, but they should be weaker on the dayside. In the Southern Hemisphere, the variation is reversed. These differences arise because of a shift in the Sun-Earth direction of the polar magnetic field line footpoints, due to a perturbation in the magnetosphere in the $x$ direction. That shift compresses or expands the circulation patterns in the ionosphere, leading to stronger and weaker currents, respectively.

Figure 6 is produced to highlight the variations predicted by Shue et al. (2002), if they exist. It shows the peak upward (positive) and downward (negative) currents, from Figures 1 and 3, as a function of magnetic local time. Currents corresponding to $B_x$ positive conditions are shown as blue solid curves and $B_x$ negative as dashed curves. If the predictions are true, there should be clear differences between the solid and dashed lines in Figure 6, and the difference should change sign once in each quadrant. The asymmetry should also be opposite between Northern and Southern Hemispheres. This behavior is not seen. Instead, we see that the peak currents during different directions of $B_x$ have remarkably similar variations with magnetic local time. Consequently, we conclude that the effect predicted by Shue et al. (2002) is either very weak or nonexistent.

## 3. Discussion and Conclusions

Our analysis shows that the IMF $B_x$ component has very little or no influence on ionospheric currents when the IMF $B_z$ is southward. In arriving at these results, we have taken great care to constrain the $B_y$ component, which otherwise would have contaminated the results because of the Parker spiral configuration of the IMF. We have also utilized three different data sets and divided those data sets into disjoint sets, which gives several independent assessments of the $B_x$ influence. In the parametrized current model (section 2.1) and for the average AMPERE maps (section 2.2) we used iterative schemes to reduce the effect of outliers, although similar conclusions are also reached using less sophisticated techniques based on simple mean values. Despite the constraints on $B_y$, and use of robust statistics, potential biases remain as follows: (1) For both Swarm/CHAMP and AMPERE, there may be differences in dipole tilt angle distributions within the data bins, and (2) for AMPERE, the solar wind-magnetosphere coupling may be systematically different in the different bins. The latter is not a concern for the CHAMP/Swarm results, since any bias should be handled by the $\epsilon$ parameter in equation (2). We address these concerns in Appendix A and show that the mentioned biases do not change the conclusions of the paper.

In the rest of this section, we briefly review previous literature on the topic of $B_x$ influence on polar ionospheric electrodynamics in the context of the present findings (sections 3.1 and 3.2). Then, in section 3.3, we present an alternative interpretation of the result, which does not involve interhemispheric asymmetries but still explains the findings by Reistad et al. (2014) and most of the $B_x$-related differences reported in the previous sections.

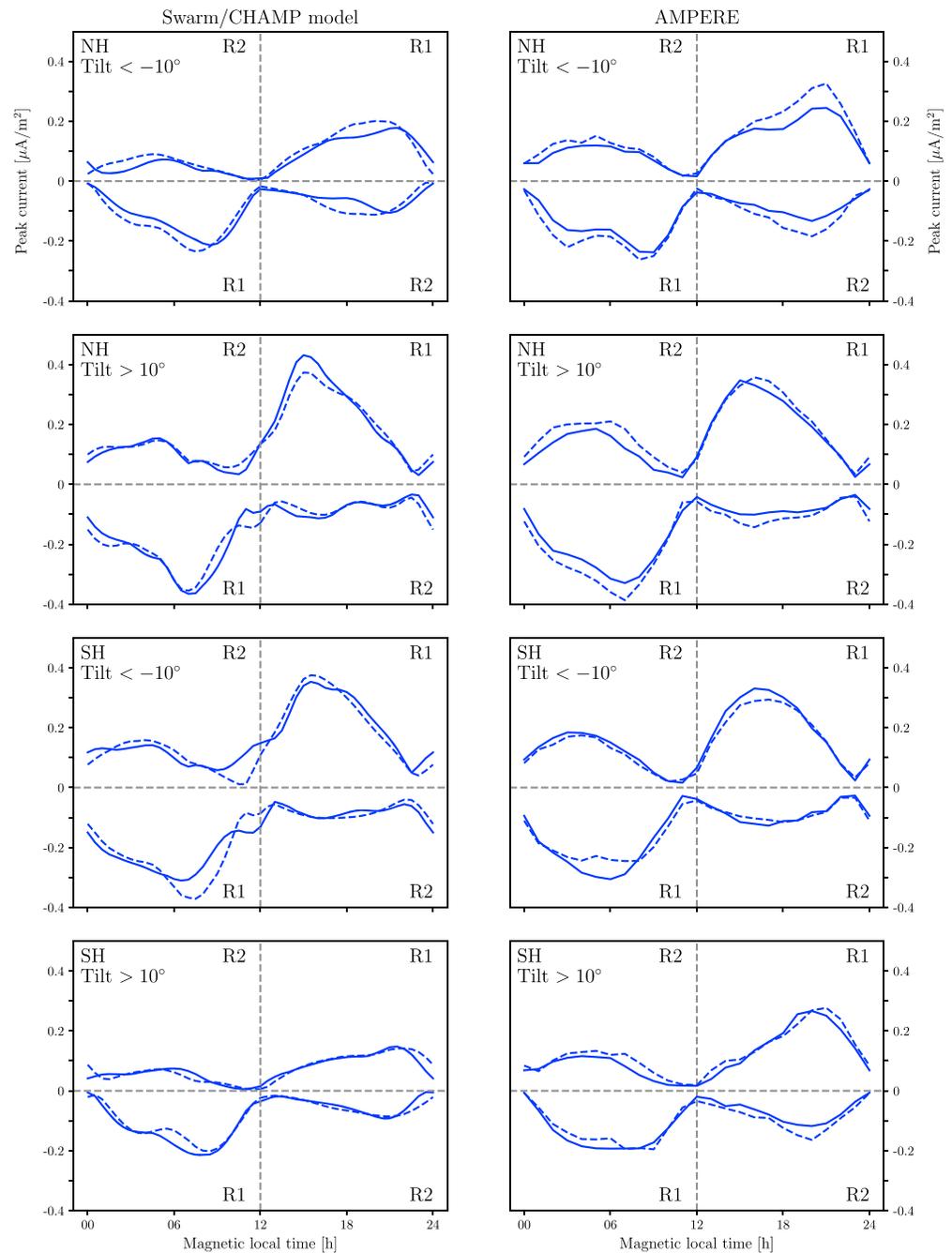

**Figure 6.** Peak upward and downward currents (positive and negative, respectively) as a function of magnetic local time. Blue solid curves show results for positive $B_x$ and dashed for negative. The values in the left column correspond to the plots in Figure 1, which shows maps based on the Swarm/CHAMP models. The values in the right column correspond to the plots in Figure 3, based on Active Magnetosphere and Planetary Electrodynamics Response Experiment (AMPERE). The four quadrants are labeled by the current that contributes to the values there (R1 or R2). Tilt angle domain and hemisphere are indicated in the top left corners. NH = Northern Hemisphere; SH = Northern Hemisphere.

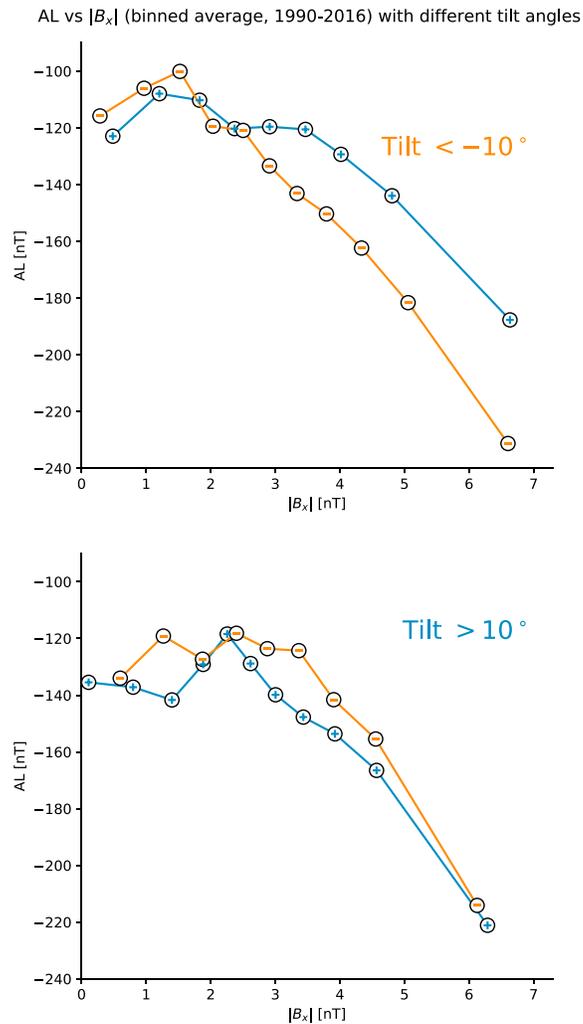

**Figure 7.** Average *AL* as a function of average $|B_x|$, in bins defined by quantiles of $B_x$ during conditions defined by $B_z < -1$ nT, $|B_y| < 1$ nT, and dipole tilt $< -10°$ (top) or $> 10°$ (bottom). In each plot, the two curves correspond to different signs of $B_x$, with blue color indicating positive $B_x$ and orange color indicating negative $B_x$. *AL* tends to be stronger (more negative) when $B_x$ and dipole tilt have the same sign.

### 3.1. Relation to Previous Studies

The observed small variation with $B_x$ is consistent with the only study that we are aware of that looks for $B_x$ effects in ionospheric convection: Förster et al. (2011) used Cluster observations, mapped to the ionosphere along model magnetic field lines, to derive global maps. By binning their data according to IMF orientation, they showed that $B_x$ had very little influence when $B_z < 0$. Interestingly, there did seem to exist a $B_x$ dependence when $B_z$ was northward. That is consistent with auroral observations by Elphinstone et al. (1990) and by Østgaard et al. (2003), who argued that $B_x$ changes magnetic field geometries such that lobe reconnection rates are not balanced between hemispheres. Lobe reconnection is believed to be minimal in the data used in the present study, since we require $B_z$ to be negative and $B_y$ small.

Our results also appear to be consistent with the numerical simulations by Peng et al. (2010), which showed no hemispheric asymmetries in ionospheric electric potential as $B_x$ increased. Only in the outer magnetosphere did $B_x$ lead to north-south asymmetries. The magnetopause position, bow shock position, and reconnection site all became asymmetrical between hemispheres, although only for the relatively uncommon case of low solar wind Alfven Mach number. Under such conditions, Peng et al. (2010) also found that increasing $B_x$ leads to reductions in cross-polar cap potential in both hemispheres. In section 3.3 we discuss how this finding can be reconciled with our results.

Distribution of $B_x$ during substorm onsets ($|B_y| < 1$, $B_z < -1$)

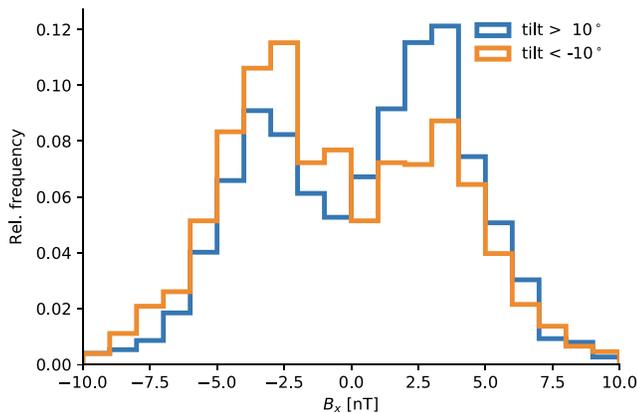

**Figure 8.** The distributions of interplanetary magnetic field $B_x$ at substorm onsets that occurred when $|B_y| < 1$ nT, $B_z < -1$ nT, and the dipole tilt angle was either $> 10°$ (blue) or $< -10°$ (orange). Each histogram is based on slightly more than 1,500 onsets, from a list of more than 62,000 substorms provided by SuperMAG (1981–2015).

### 3.2. Asymmetries in the Aurora Associated With $B_x$

Taken at face value, our results seem to contradict the findings by Shue et al. (2002) and Reistad et al. (2014), who found that $B_x$ has a small but significant influence on auroral intensity. This influence was explained in terms of ionospheric currents, controlled by $B_x$ through mechanisms analogous to how $B_y$ affects the magnetosphere (Cowley, 1981).

Because of this discrepancy, we should consider the possibility that the observations by Shue et al. (2002) and Reistad et al. (2014) of $B_x$-dependent asymmetries do not reflect a true $B_x$ control, due to observational and/or technical issues. While we are not able to point out any specific problems in these studies, this conclusion cannot be ruled out. It is well known that the space-based UV imagers that were used are prone to variations that are difficult to fully account for in statistical studies and which may lead to hidden biases, especially the results by Shue et al. (2002) from the Northern Hemisphere summer season that might be questioned, due to the impact of sunlight on UV auroral images and the difficulty in correcting for this contamination. As we will show below, if we disregard this last observation, the auroral observations can be reconciled with most of the current observations in the present paper if the interpretation is changed. However, considering the small magnitudes of the observed $B_x$ variations, we suggest testing the auroral observations with other independent data sets, for example, the particle and/or optical instrumentation on board the Defense Meteorological Satellite Program (DMSP) satellites.

A second possible explanation for the discrepancy may have to do with a seasonal difference in the magnitude of the $B_x$ asymmetries. In the premidnight region, there seems to be a nonlinear relationship between current and aurora, where increasing auroral intensity ionizes the atmosphere, leading to increasing currents, which in turn increases the aurora (see, e.g., Ohtani et al., 2009). This amplifying effect in darkness may be the reason why only 2 out of 11 comparisons from the winter hemisphere in Figure 5 are inconsistent with the Reistad et al. (2014) predictions. However, it is puzzling that the CHAMP/Swarm results consistently show an opposite asymmetry of similar magnitude during local summer. In the next section we present a new interpretation which explains this, as well as additional evidence which is more in line with this new idea rather than nightside amplification.

### 3.3. Increased Geoefficiency When $B_x$ and Dipole Tilt Have the Same Sign

A third interpretation of the discrepancy between the auroral observations and the observations in the present paper is possible: That the observations by Shue et al. (2002) and Reistad et al. (2014) about $B_x$-dependent auroras reflect a real difference, but their interpretation in terms of hemispherically asymmetric currents is wrong. This conclusion calls for an alternative mechanism. Here we propose that the dominating effect of $B_x$ in this and the previous studies is to change the overall coupling efficiency between the solar wind and the magnetosphere, allowing more energy to be transferred when $B_x$ and the dipole tilt have the same sign.

We suggest that the mechanism behind this effect is related to how the dipole tilt angle and $B_x$ affect the location of the dayside reconnection site: Models (Hoilijoki et al., 2014; Park et al., 2006; Russell et al., 2003) and observations (Zhu et al., 2015) show that the dayside reconnection site shifts southward (northward) of the subsolar point when the dipole tilt angle is positive (negative). That is, the reconnection site tends to follow the Earth's equatorial plane as it tilts away from the Sun-Earth line. Hoilijoki et al. (2014) showed that $B_x$ can equalize this shift, if it has the same sign as the tilt angle. Under such conditions, the reconnection site will be located near the subsolar point, where it is most efficient (Park et al., 2006).

This idea is consistent with the results by Reistad et al. (2014), who observed stronger aurora (in the north) when $B_x$ and dipole tilt was negative, compared to $B_x$ positive conditions, and stronger aurora (in the south) when $B_x$ and dipole tilt was positive, compared to $B_x$ negative conditions. As mentioned before, this can be interpreted both as an interhemispheric asymmetry imposed by $B_x$ which is independent of seasons or, as we do here, an influence of $B_x$ on the overall geoefficiency, which changes with the sign of the dipole tilt angle. Shue et al. (2002) reported $B_x$ asymmetries from the Northern Hemisphere during both winter and summer.

The new idea is only consistent with their winter results, but as mentioned above, there are reasons to place less emphasis on the summer results due to the contamination of sunlight on the images.

This idea is also consistent with all the Birkeland current asymmetries observed in the CHAMP/Swarm models. It is also consistent with most of the horizontal current maps, from CHAMP/Swarm and SuperMAG: Only one out of six shows an inconsistent asymmetry, and one shows no asymmetry. The main discrepancy with this interpretation is found in the AMPERE comparisons, with only two out of eight comparisons in agreement. There may, however, be reasons to place less emphasis on the AMPERE comparisons than the CHAMP/Swarm results. The AMPERE currents are derived from magnetometers that are much less accurate than those flying on CHAMP and Swarm, a shortcoming which is not necessarily compensated by the much greater number of spacecraft. Furthermore, the distribution of orbits in magnetic coordinates may create complications in the current estimates which affect the results, especially in the Southern Hemisphere (Anderson et al., 2017), where none of the comparisons happen to fit the new explanation.

If the solar wind-magnetosphere coupling is stronger when dipole tilt and $B_x$ are in the same direction, other indicators of geomagnetic activity should be affected as well. In Figures 7 and 8 we test this concept on the $AL$ index and on substorm occurrence, respectively.

Figure 7 shows binned averages of $AL$, plotted against binned average $B_x$. The bins are defined by $|B_y| < 1$, $B_z < -1$; dipole tilt angle either $< -10°$ (top) or $> 10°$ (bottom); and a set of quantiles for $B_x$ during these conditions. That means that the bins contain an almost equal number of samples but that their $B_x$ boundaries are variable. The averages are computed robustly, with iterative reweighting by Huber weights. The x axis shows the absolute value of $B_x$, and corresponding values of $AL$ are labeled by + or − (blue and orange, respectively), depending on the sign of $B_x$. The figure clearly shows that $AL$ tends to be stronger if $B_x$ and the tilt angle have the same sign. This is consistent in all bins beyond $|B_x| \approx 2.5nT$. Each bin is based on independent datapoints, with more than >12,000 samples per point in both panels. The data set, from the OMNI database, covers the period 1990–2016.

The difference does not seem to increase with increasing $B_x$, and it even decreases slightly with increasing $B_x$ when the tilt angle is positive. It is possible that the effect maximizes for some angle between the IMF in the $xz$ plane and the dipole axis. It may be that larger $B_x$ overcompensates for the shift in reconnection region associated with the dipole tilt and thus reduces the coupling efficiency. We note that this behavior is different from what we would expect from the Reistad et al. (2014) and Shue et al. (2002) mechanisms, which is that increasing $B_x$ increases asymmetries.

If $B_x$ affects the dayside reconnection rate in the manner proposed here, we should expect more substorms to occur when $B_x$ and the tilt angle have the same sign. This is because more magnetic flux will be opened, and more substorms are therefore required to close it. We test this prediction in Figure 8, which is based on the SuperMAG substorm list (Newell & Gjerloev, 2011). It shows the distribution of $B_x$ during substorm onsets that happened between 1981 and 2016, when $|B_y| < 1$ nT, $B_z < -1$ nT, and the dipole tilt was either $< -10°$ (orange) or $> 10°$ (blue). Each distribution is based on more than 1,500 substorms. The distributions are clearly asymmetrical, with positive $B_x$ favored when the dipole tilt was positive and negative $B_x$ when the tilt angle was negative. This result is also different from what we would expect from the Reistad et al. (2014) and Shue et al. (2002) mechanisms, which do not involve any change in overall solar wind coupling efficiency. The result supports the idea that the dayside reconnection rate is stronger when the IMF $B_x$ and the dipole tilt angle have the same signs than when they are opposite.

## 4. Conclusions

We have shown through various observations that $B_x$ leads to very small or negligible interhemispheric differences in ionospheric currents. This contradicts interpretations made in previous studies, to explain observations of $B_x$-dependent asymmetries in the aurora.

While our results contradict previous predictions about interhemispheric differences associated with $B_x$, they do suggest a small and seasonally dependent influence of $B_x$ on the efficiency of the solar wind-magnetosphere coupling. This coupling tends to be stronger when $B_x$ and the dipole tilt angle have the same sign. Under such conditions, the subsolar reconnection site is closer to the subsolar point (e.g., Hoilijoki et al., 2014) and presumably more efficient.

The idea of a $B_x$-dependent coupling efficiency is supported by observations of average $AL$ magnitudes and of substorm occurrence. However, contradictory observations are also found, notably with AMPERE and in the observations by Shue et al. (2002) in the summer hemisphere. There is therefore a need for further investigations on this topic, with independent data sets, and on different quantities, such as ionospheric convection. There is also a need for a more detailed investigation of how the solar wind Alfven Mach number might change the $B_x$ effect; the modeling by Peng et al. (2010) suggests that it becomes less important when the Mach number is large.

The $B_x$ effect observed here is not, as far as we know, part of any solar wind-magnetosphere coupling function (e.g., Newell et al., 2007; Tenfjord & Østgaard, 2013, and references therein). Our findings suggest that significant improvements can be made by taking it into account.

## Appendix A: Potential Biases in the Data Sets

Figure A1 shows the distribution of the Newell et al. (2007) coupling function, $\epsilon$, across the maps used to produce the average currents of Figure 3. Significantly different distributions are expected to be associated with different current strengths, due to differences in solar wind-magnetosphere energy transfer. This could potentially obscure any $B_x$ effect. The figure shows that the distributions are not identical, with more events having large values of $\epsilon$ when $B_x > 2$ nT and the dipole tilt angle (denoted by $\beta$ in the figure) was $> 10°$. The observed differences with respect to $B_x$ (Figure 3) are in the opposite direction, however, which suggests that this bias is not a major influence.

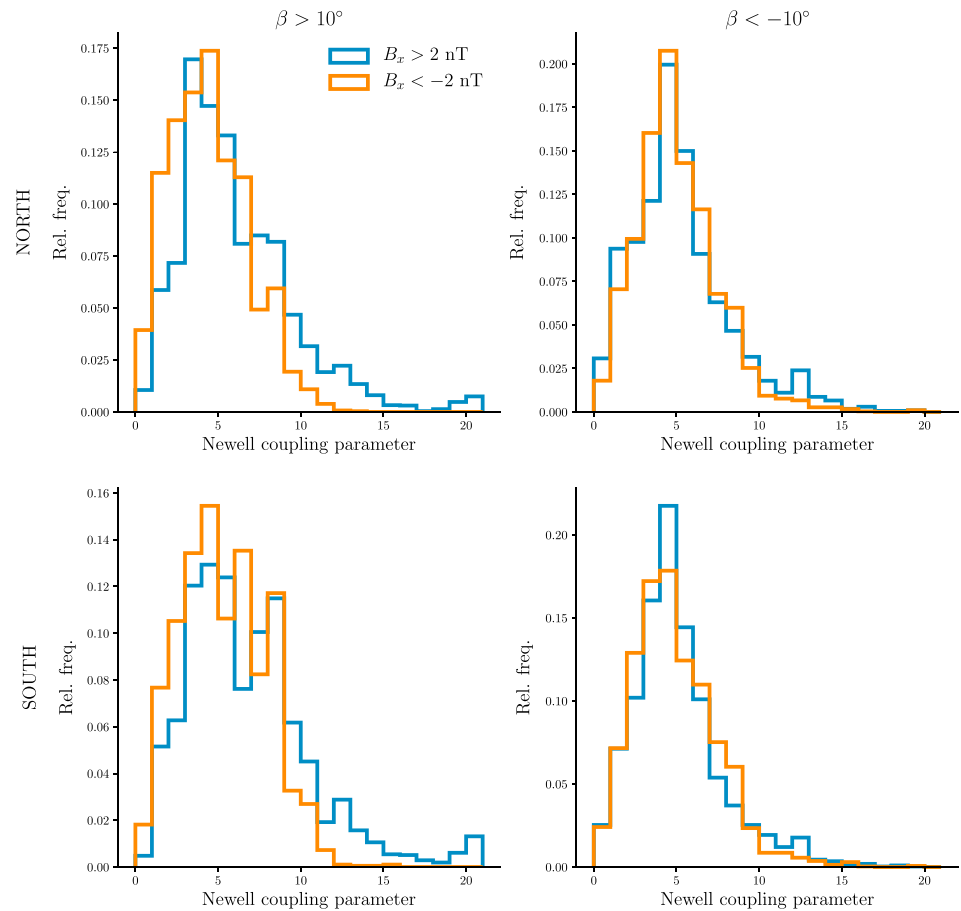

**Figure A1.** The distribution of Newell et al.'s, 2007 coupling parameter ($\epsilon$ in equation (2) for AMPERE current maps used to make Figure (3). Each plot corresponds to one hemisphere (top row: North hemisphere, bottom row: South hemisphere) and tilt angle (left column: positive tilt, right column: negative tilt), and the two histograms in each plot correspond to different $B_x$ bins (blue: $B_x > 2$ nT, orange: $B_x < -2$ nT).

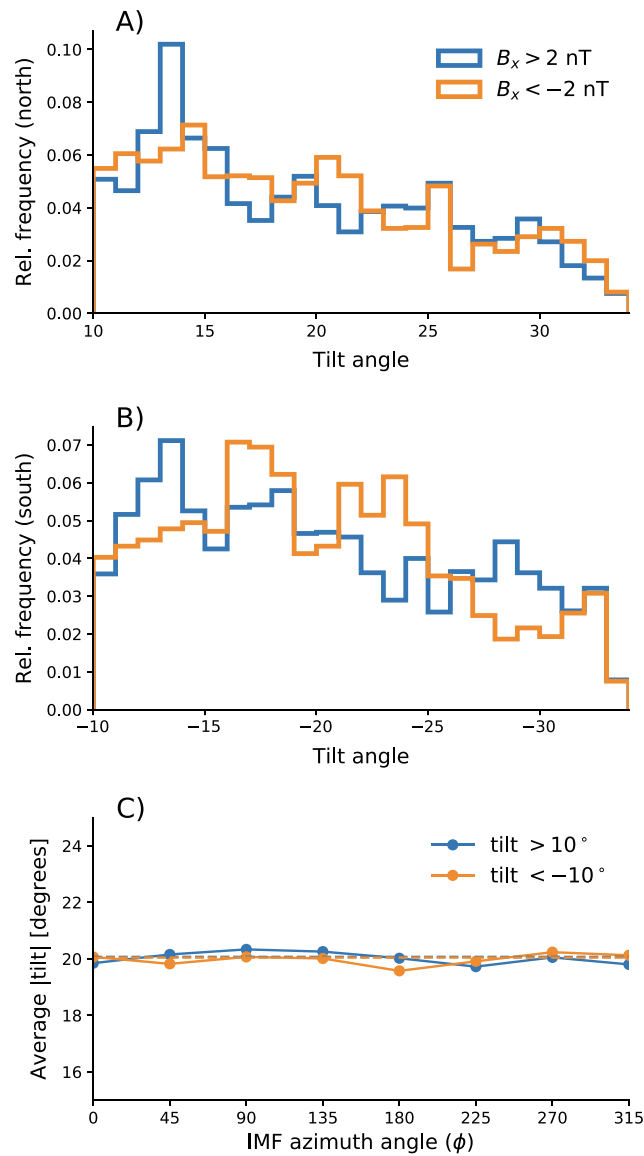

**Figure A2.** (a, b) Distributions of the dipole tilt angle at the times of AMPERE maps that were used to produce Figure 3. Panel (a) corresponds to maps in the Northern Hemisphere during positive tilt and panel (b) to maps in the Southern Hemisphere during negative tilt. Different colors correspond to different signs of $B_x$. (c) The weighted mean tilt angle in 45° wide bins of interplanetary magnetic field azimuth angle, at the times of the CHAMP/Swarm data used to produce Figure 1. The weights used to calculate the means are the final Huber weights in the iterative scheme to estimate the spherical harmonic models described in section 2.1.

Figure A2 is produced to reveal any biases in dipole tilt angle distributions that might influence the comparisons of different $B_x$ conditions using Figures 1 and 3. Figures A2a and A2b show the distributions of the dipole tilt angle at the times of the AMPERE maps used to make Figures 3c and 3d (3e and 3f). The distributions in Figure A2a are quite similar, showing that comparisons between Figures 3c and 3d are not influenced by tilt angle. The distributions in panel b are slightly skewed toward more sunlit conditions, and thus higher conductivities, for periods when $B_x > 2$ nT compared to $B_x < -2$ nT. This small bias possibly enhances the apparent $B_x$ effect suggested by comparing Figures 3e and 3f. Figure A2c shows the weighted mean absolute tilt angle, in 45° wide bins of IMF azimuth angle, $\phi$, at the times of the CHAMP/Swarm data points that were used to produce Figure 1. The weights are the Huber weights of the last iteration in the iterative scheme used to estimate the model parameters. The essentially flat curves show that the results of the analysis in section 2.1 are not influenced by biases in dipole tilt angle distributions.

## Appendix B: $B_y$ Effects

This paper is mainly concerned with $B_x$ effects. Since there is a significant anticorrelation between $B_x$ and $B_y$, we have had to control for $B_y$, by either constraining its magnitude or by parametrization. In this section we present figures on a similar format as in sections 2.1, 2.2, and 2.3, only with the roles of $B_x$ and $B_y$ reversed. The purpose is to demonstrate that the techniques are capable of reproducing known variations with $B_y$, reported in numerous studies (e.g., Anderson et al., 2008; Friis-Christensen & Wilhjelm, 1975; Friis-Christensen et al., 1984; Green et al., 2009; Laundal, Gjerloev, et al., 2016; Papitashvili et al., 2002; Weimer, 2001).

Figure B1 shows the $B_y$ effect according to the empirical models based on CHAMP and Swarm (section 2.1). Figure B2 shows the $B_y$ effect according to AMPERE (see section 2.2 for details) and Figure B3 shows $B_y$ effects according to the analysis based on SuperMAG ground magnetometers (see section 2.3).

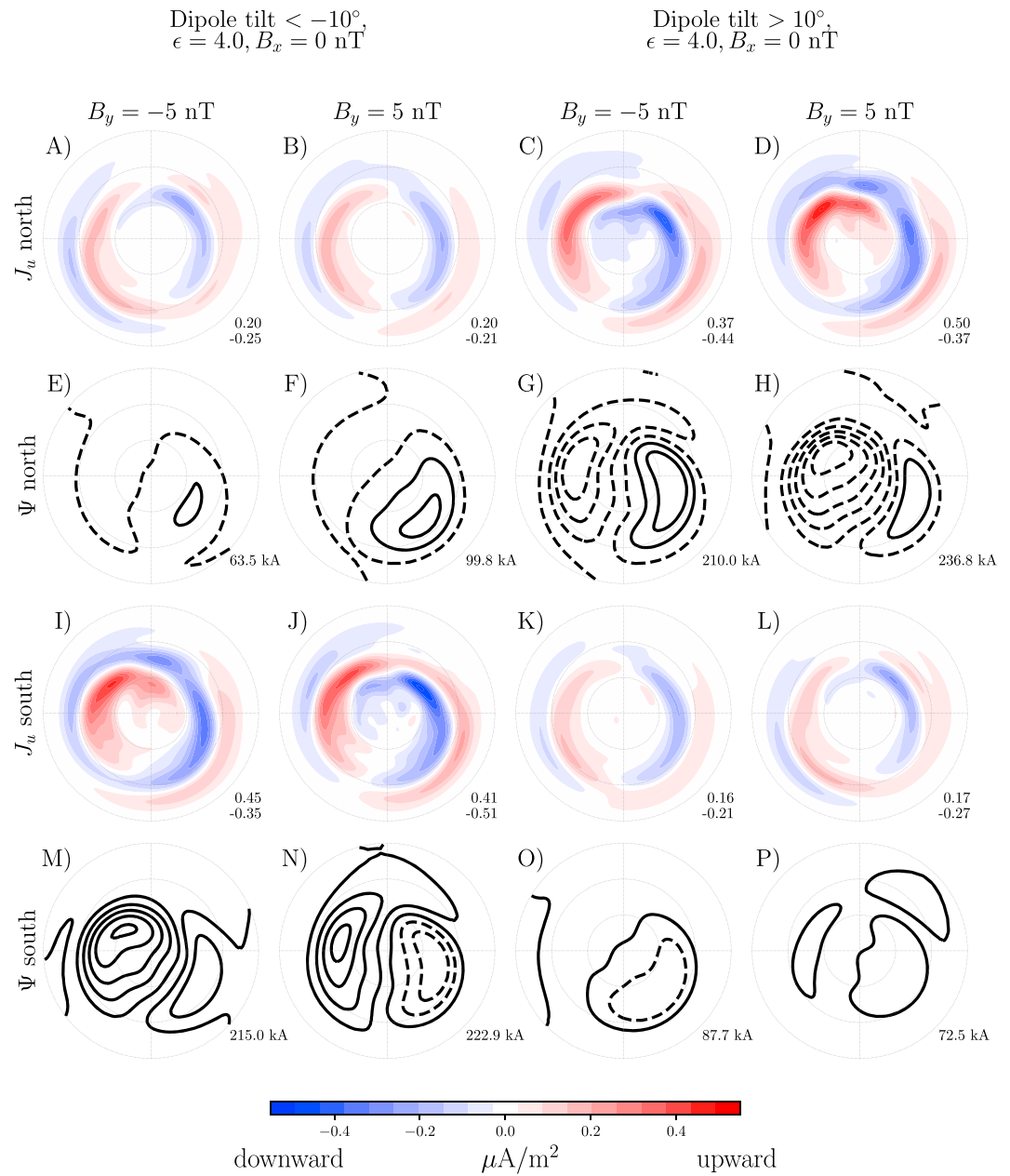

**Figure B1.** Same as Figure 1 but with the roles of $B_x$ and $B_y$ reversed.

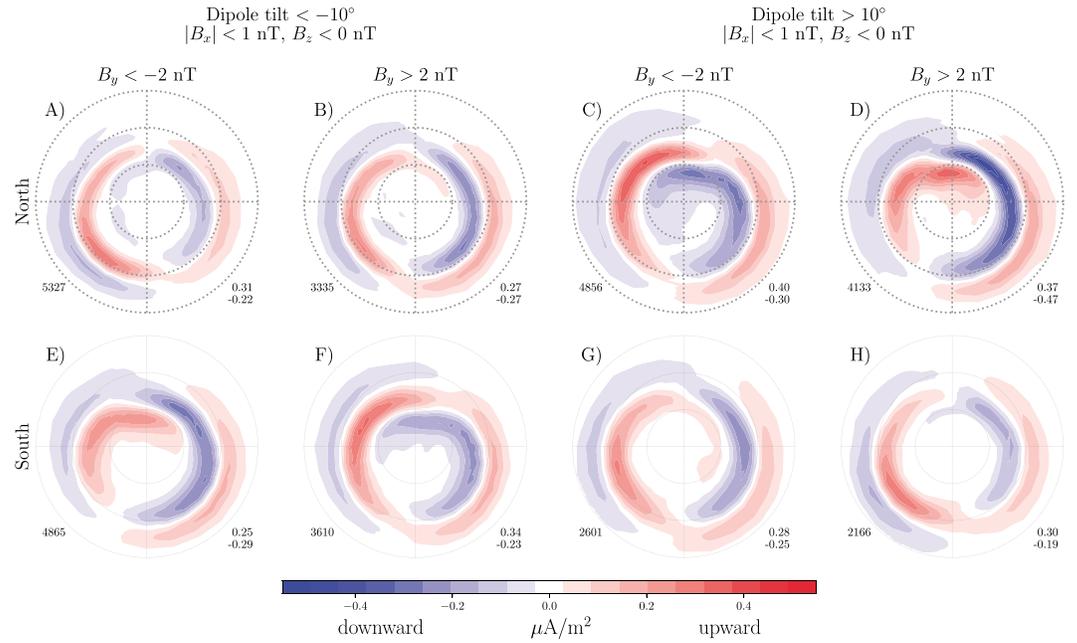

**Figure B2.** Same as Figure 3 but with the roles of $B_x$ and $B_y$ reversed.

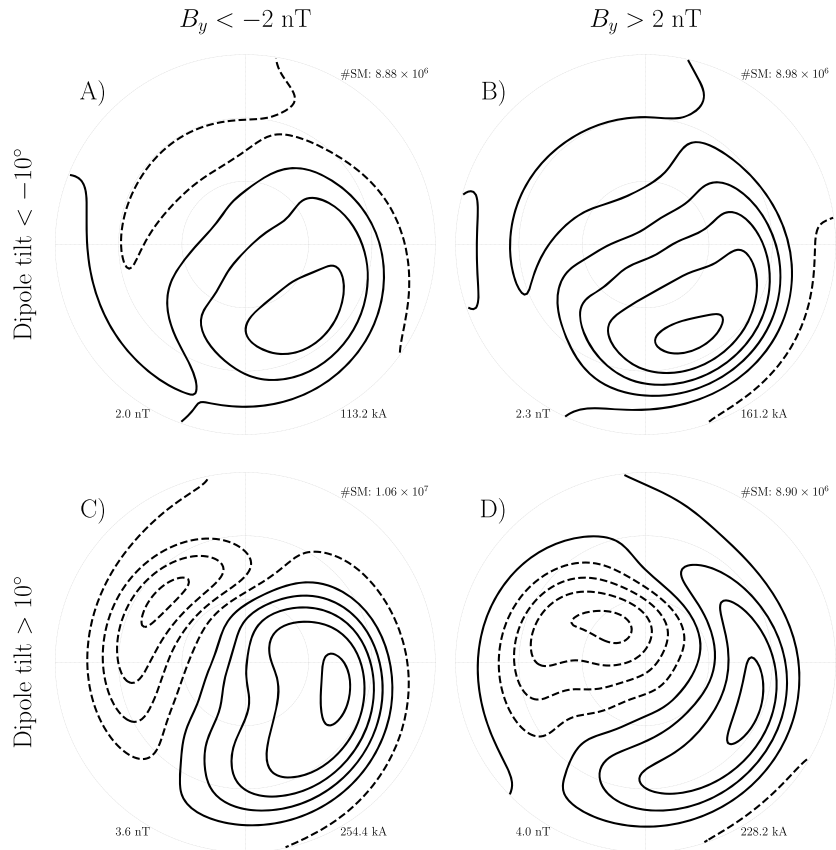

**Figure B3.** Same as Figure 4 but only with the roles of $B_x$ and $B_y$ reversed.

**Acknowledgments**

The support of the CHAMP mission by the German Aerospace Center (DLR) and the Federal Ministry of Education and Research is gratefully acknowledged. CHAMP magnetic field measurements are available through the Information System and Data Center, ISDC, at http://isdc.gfz-potsdam.de. We thank the European Space Agency for providing Swarm magnetic field measurements, accessible at http://earth.esa.int/swarm. We thank the AMPERE team and the AMPERE Science Center for providing the Iridium-derived data products, accessible at http://ampere.jhuapl.edu/. For the ground magnetometer data, downloaded from http://supermag.jhuapl.edu/, we gratefully acknowledge the SuperMAG data providers and P.I. Jesper Gjerloev. We acknowledge use of NASA/GSFC's Space Physics Data Facility's OMNIWeb service (https://omniweb.gsfc.nasa.gov/), and OMNI data. This data set includes the $AL$ index and time-shifted solar wind and interplanetary magnetic field measurements. This study was supported by the Research Council of Norway/CoE under contract 223252/F50.

# References

Anderson, B. J., Korth, H., Waters, C. L., Green, D. L., & Stauning, P. (2008). Statistical Birkeland current distributions from magnetic field observations by the iridium constellation. *Annals of Geophysics*, 26, 671–687.

Anderson, B. J., Korth, H., Welling, D. T., Merkin, V. G., Wiltberger, M. J., Raeder, J., et al. (2017). Comparison of predictive estimates of high-latitude electrodynamics with observations of global-scale Birkeland currents. *Space Weather*, 15, 352–373. https://doi.org/10.1002/2016SW001529

Anderson, B. J., Takahashi, K., & Toth, B. A. (2000). Sensing global Birkeland currents with Iridium engineering magnetometer data. *Geophysical Research Letters*, 27, 4045–4048.

Anderson, E., Bai, Z., Bischof, C., Blackford, S., Demmel, J., Dongarra, J., et al. (1999). *LAPACK users' guide* (3rd ed.). Philadelphia, PA: Society for Industrial and Applied Mathematics.

Backus, G. (1986). Poloidal and toroidal fields in geomagnetic field modeling. *Reviews of Geophysics*, 24, 75–109.

Baker, J. B., Ridley, A. J., Papitashvili, V. O., & Clauer, C. R. (2003). The dependence of winter aurora on interplanetary parameters. *Journal of Geophysical Research*, 108(A4), 8009. https://doi.org/10.1029/2002JA009352

Cnossen, I., Richmond, A. D., & Wiltberger, M. (2012). The dependence of the coupled magnetosphere-ionosphere-thermosphere system on the Earth's magnetic dipole moment. *Journal of Geophysical Research*, 117, A05302. https://doi.org/10.1029/2012JA017555

Cowley, S. W. (1981). Asymmetry effects associated with the x-component of the IMF in a magnetically open magnetosphere. *Planetary and Space Science*, 29(8), 809–818.

Elphinstone, R. D., Jankowska, K., Murphree, J. S., & Cogger, L. L. (1990). The configuration of the auroral distribution for interplanetary magnetic field $B_z$ northward 1. IMF $B_x$ and $B_y$ dependencies as observed by the viking satellite. *Journal of Geophysical Research*, 95, 5791–5804.

Finlay, C. C., Olsen, N., Kotsiaros, S., Gillet, N., & Tøffner-clausen, L. (2016). Recent geomagnetic secular variation from Swarm and ground observatories as estimated in the CHAOS-6 geomagnetic field model, Earth. *Planets and Space*, 68, 112. https://doi.org/10.1186/s40623-016-0486-1

Förster, M., Feldstein, Y. I., Gromova, L. I., Dremukhina, L. A., Levitin, A. E., & Haaland, S. E. (2011). Plasma convection in the high-latitude ionosphere deduced from Cluster EDI data and IMF Bx component. In *Proc. 34th Apatity Annual Seminar "Physics of Auroral Phenomena", Apatity, 2011* (pp. 42–45).

Friis-Christensen, E., Kamide, Y., Richmond, A. D., & Matsushita, S. (1984). Interplanetary magnetic field control of high-latitude electric fields and currents determined from Greenland magnetometer data. *Journal of Geophysical Research*, 90, 1325–1338.

Friis-Christensen, E., & Wilhjelm, J. (1975). Polar cap currents for different directions of the interplanetary magnetic field in the *Y-Z* plane. *Journal of Geophysical Research*, 80, 1248–1260. https://doi.org/10.1029/JA080i010p01248

Gjerloev, J. W. (2012). The superMAG data processing technique. *Journal of Geophysical Research*, 117, A09213. https://doi.org/10.1029/2012JA017683

Green, D. L., Waters, C. L., Anderson, B. J., & Korth, H. (2009). Seasonal and interplanetary magnetic field dependence of the field-aligned currents for both Northern and Southern Hemispheres. *Annals of Geophysics*, 27, 1701–1715. https://doi.org/10.5194/angeo-27-1701-2009

Haaland, S. E., Paschmann, G., Förster, M., Quinn, J. M., Torbert, R. B., McIlwain, C. E., et al. (2007). High-latitude plasma convection from Cluster EDI measurements: Method and IMF-dependence. *Annals of Geophysics*, 25, 239–253. https://doi.org/10.5194/angeo-25-239-2007

Heppner, J. P., & Maynard, N. C. (1987). Empirical high-latitude electric field models. *Journal of Geophysical Research*, 92, 4467–4489. https://doi.org/10.1029/JA092iA05p04467

Hoilijoki, S., Souza, V. M., Walsh, B. M., Janhunen, P., & Palmroth, M. (2014). Magnetopause reconnection and energy conversion as influenced by the dipole tilt and IMF $b_x$. *Journal of Geophysical Research*, 119, 4484–4494. https://doi.org/10.1002/2013JA019693

Huber, P. J. (1964). Robust estimation of location parameter. *Annals of Mathematical Statistics*, 35, 73–101.

Iijima, T., & Potemra, T. A. (1978). Large-scale characteristics of field-aligned currents associated with substorms. *Journal of Geophysical Research*, 83, 599–615.

Laundal, K. M., Cnossen, I., Milan, S. E., Haaland, S. E., Coxon, J., Pedatella, N. M., et al. (2017). North-south asymmetries in Earth's magnetic field: Effects on high-latitude geospace. *Space Science Reviews*, 206, 225–257. https://doi.org/10.1007/s11214-016-0273-0

Laundal, K. M., Finlay, C. C., & Olsen, N. (2016). Sunlight effects on the 3D polar current system determined from low Earth orbit measurements. *Earth Planets Space*, 68, 142. https://doi.org/10.1186/s40623-016-0518-x

Laundal, K. M., & Gjerloev, J. W. (2014). What is the appropriate coordinate system for magnetometer data when analyzing ionospheric currents? *Journal of Geophysical Research*, 119, 8637–8647. https://doi.org/10.1002/2014JA020484

Laundal, K. M., Gjerloev, J. W., Ostgaard, N., Reistad, J. P., Haaland, S. E., Snekvik, K., et al. (2016). The impact of sunlight on high-latitude equivalent currents. *Journal of Geophysical Research*, 121, 2715–2726. https://doi.org/10.1002/2015JA022236

Laundal, K. M., & Østgaard, N. (2009). Asymmetric auroral intensities in the Earth's Northern and Southern Hemispheres. *Nature*, 460, 491–493. https://doi.org/10.1038/nature08154

Laundal, K. M., & Richmond, A. D. (2017). Magnetic coordinate systems. *Space Science Reviews*, 206, 27–59. https://doi.org/10.1007/s11214-016-0275-y

Matsuo, T., Knipp, D. J., Richmond, A. D., Kilcommons, L., & Anderson, B. J. (2015). Inverse procedure for high-latitude ionospheric electrodynamics: Analysis of satellite-borne magnetometer data. *Journal of Geophysical Research*, 120, 5241–5251. https://doi.org/10.1002/2014JA020565

Newell, P. T., & Gjerloev, J. W. (2011). Evaluation of superMAG auroral electrojet indices as indicators of substorms and auroral power. *Journal of Geophysical Research*, 116, A12211. https://doi.org/10.1029/2011JA016779

Newell, P. T., Ruohoniemi, J. M., & Meng, C.-I. (2004). Maps of precipitation by source region, binned by IMF, with inertial convection streamlines. *Journal of Geophysical Research*, 109, A10206. https://doi.org/10.1029/2004JA010499

Newell, P. T., Sotirelis, T., Liou, K., Meng, C. I., & Rich, F. J. (2007). A nearly universal solar wind-magnetosphere coupling function inferred from 10 magnetospheric state variables. *Journal of Geophysical Research*, 112, A01206. https://doi.org/10.1029/2006JA012015

Ohtani, S., Wing, S., Ueno, G., & Higuchi, T. (2009). Dependence of premidnight field-aligned currents and particle precipitation on solar illumination. *Journal of Geophysical Research*, 114, A12205. https://doi.org/10.1029/2009JA014115

Olsen, N. (1997). Ionospheric $F$ region currents at middle and low latitudes estimated from Magsat data. *Journal of Geophysical Research*, 102, 4563–4576.

Østgaard, N., Mende, S. B., Frey, H. U., Frank, L. A., & Sigwarth, J. B. (2003). Observations of non-conjugate theta aurora. *Geophysical Research Letters*, 30, 2125. https://doi.org/10.1029/2003GL017914

Papitashvili, V. O., Christiansen, F., & Neubert, T. (2002). A new model of field-aligned currents derived from high-precision satellite magnetic field data. *Geophysical Research Letters*, *29*, 2125. https://doi.org/10.1029/2001GL014207

Park, K. S., Ogino, T., & Walker, R. J. (2006). On the importance of antiparallel reconnection when the dipole tilt and IMF $B_y$ are nonzero. *Journal of Geophysical Research*, *111*, A05202. https://doi.org/10.1029/2004JA010972

Parker, E. N. (1958). Dynamics of the interplanetary gas and magnetic fields. *Astrophysical Journal*, *128*, 664–676. https://doi.org/10.1086/146579

Peng, Z., Wang, C., & Hu, Y. Q. (2010). Role of IMF $B_x$ in the solar wind-magnetosphere-ionosphere coupling. *Journal of Geophysical Research*, *115*, A08224. https://doi.org/10.1029/2010JA015454

Pettigrew, E. D., Shepherd, S. G., & Ruohoniemi, J. M. (2010). Climatological patterns of high-latitude convection in the Northern and Southern hemispheres: Dipole tilt dependencies and interhemispheric comparison. *Journal of Geophysical Research*, *115*, A07305. https://doi.org/10.1029/2009JA014956

Reistad, J. P. Østgaard, N., Laundal, K. M., Haaland, S., Tenfjord, P., Snekvik, K., et al. (2014). Intensity asymmetries in the dusk sector of the poleward auroral oval due to IMF $B_x$. *Journal of Geophysical Research*, *119*, 9497–9507. https://doi.org/10.1002/2014JA020216

Reistad, J. P. Østgaard, N., Laundal, K. M., & Oksavik, K. (2013). On the non-conjugacy of nightside aurora and their generator mechanisms. *Journal of Geophysical Research*, *118*, 3394–3406. https://doi.org/10.1002/jgra.50300

Richmond, A. D. (1995a). Ionospheric electrodynamics using magnetic apex coordinates. *Journal of Geomagnetism and Geoelectricity*, *47*, 191–212.

Richmond, A. D. (1995b). Ionospheric electrodynamics. In A. Richmond (Ed.), *Handbook of atmospheric electrodynamics, volume II* (pp. 249–290). CRC Press.

Russell, C. T., & McPherron, R. L. (1973). Semiannual variation of geomagnetic activity. *Journal of Geophysical Research*, *78*(1), 92–108. https://doi.org/10.1029/JA078i001p00092

Russell, C. T., Wang, Y. L., & Raeder, J. (2003). Possible dipole tilt dependence of dayside magnetopause reconnection. *Geophysical Research Letters*, *30*, 1937. https://doi.org/10.1029/2003GL017725

Sabaka, T. J., Hulot, G., & Olsen, N. (2010). Handbook of geomathematics. In *Chap. mathematical properties relevant to geomagnetic field modeling* (pp. 503–538). Berlin, Heidelberg: Springer. https://doi.org/10.1007/978-3-642-01546-5_17

Shi, Q. Q., Zong, Q.-G., Fu, S. Y., Dunlop, M. W., Pu, Z. Y., Parks, G. K., et al. (2013). Solar wind entry into the high-latitude terrestrial magnetosphere during geomagnetically quiet times. *Nature Communications*, *4*, 1466.

Shue, J.-H., Newell, P. T., Liou, K., & Meng, C. I. (2001). Influence of interplanetary magnetic field on global auroral patterns. *Journal of Geophysical Research*, *106*, 5913–5926. https://doi.org/10.1029/2000JA000325

Shue, J.-H., Newell, P. T., Liou, K., Meng, C.-I., & Cowley, S. W. H. (2002). Interplanetary magnetic field $b_x$ asymmetry effect on auroral brightness. *Journal of Geophysical Research*, *107*(A8), 1197. https://doi.org/10.1029/2001JA000229

Stenbaek-Nielsen, H. C., Wescott, E. M., Davis, T. N., & Peterson, R. W. (1973). Differences in auroral intensity at conjugate points. *Journal of Geophysical Research*, *78*, 659–671.

Tenfjord, P., & Østgaard, N. (2013). Energy transfer and flow in the solar wind-magnetosphere-ionosphere system: A new coupling function. *Journal of Geophysical Research*, *118*, 5659–5672. https://doi.org/10.1002/jgra.50545

Tenfjord, P., Østgaard, N., Snekvik, K., Laundal, K. M., Reistad, J. P., Haaland, S., & Milan, S. (2015). How the IMF $B_y$ induces a $B_y$ component in the closed magnetosphere and how it leads to asymmetric currents and convection patterns in the two hemispheres. *Journal of Geophysical Research: Space Physics*, *120*, 9368–9384. https://doi.org/10.1002/2015JA021579

Waters, C. L., Anderson, B. J., & Liou, K. (2001). Estimation of global field aligned currents using the iridium system magnetometer data. *Geophysical Research Letters*, *28*, 2165–2168. https://doi.org/10.1029/2000GL012725

Weimer, D. (2001). Maps of ionospheric field-aligned currents as a function of the interplanetary magnetic field derived from Dynamics Explorer 2. *Journal of Geophysical Research*, *106*, 12,889–12,902. https://doi.org/10.1029/2000JA000295

Weimer, D. R. (2013). An empirical model of ground-level geomagnetic perturbations. *Space Weather*, *11*, 107–120. https://doi.org/10.1002/swe.20030

Wilcox, J. M., & Ness, N. F. (1965). Quasi-stationary corotating structure in the interplanetary medium. *Journal of Geophysical Research*, *70*(23), 5793–5805. https://doi.org/10.1029/JZ070i023p05793

Zhao, H., & Zong, Q.-G. (2012). Seasonal and diurnal variation of geomagnetic activity: Russell-mcpherron effect during different imf polarity and/or extreme solar wind conditions. *Journal of Geophysical Research*, *117*, A11222. https://doi.org/10.1029/2012JA017845

Zhu, C. B., Zhang, H., Ge, Y. S., Pu, Z. Y., Liu, W. L., Wan, W. X., et al. (2015). *Dipole tilt angle effect on magnetic reconnection locations on the magnetopause* (Vol. 120, pp. 5344–5354). https://doi.org/10.1002/2015JA020989